\pdfoutput=1

\documentclass[useAMS,usenatbib,fleqn]{mnras}

\usepackage{graphicx}
\usepackage{amsmath,amssymb}
\usepackage{natbib}

\title[Orbit classification in a pseudo-Newtonian problem]{Orbit classification in a pseudo-Newtonian Copenhagen problem with Schwarzschild-like primaries}

\author[Zotos et al.]{Euaggelos E. Zotos$^1$\thanks{E-mail: evzotos@physics.auth.gr}, Fredy L. Dubeibe$^2$\thanks{E-mail: fldubeibem@unal.edu.co}, Jan Nagler$^3$\thanks{E-mail: jnagler@ethz.ch} and Emilio Tejeda$^{4,5}$\thanks{E-mail: emilio.tejeda@conacyt.mx} \\
$^1$ Department of Physics, School of Science, Aristotle University of Thessaloniki, GR-541 24, Thessaloniki, Greece \\
$^2$ Grupo de Investigaci\'{o}n Cavendish, Facultad de Ciencias Humanas y de la Educaci\'{o}n, Universidad de los Llanos, \\ Villavicencio 500017, Colombia \\
$^3$ Deep Dynamics Group \& Centre for Human and Machine Intelligence, Frankfurt School of Finance \& Management, \\ Frankfurt, Germany \\
$^4$ CONACyT -- Instituto de F\'{i}sica y Matem\'{a}ticas, Universidad Michoacana de San Nicol\'{a}s de Hidalgo, \\
     Edificio C-3, Ciudad Universitaria, 58040 Morelia, Michoac\'{a}n, Mexico \\
$^5$ Instituto de Astronom\'{i}a, Universidad Nacional Aut\'{o}noma de M\'exico, AP 70-264, 04510 Ciudad de M\'{e}xico, Mexico
}

\begin{document}

\date{Accepted 2019 May 21. Received 2019 May 21; in original form 2019 February 26}

\pubyear{2019} \volume{487} \pagerange{2340--2353}

\setcounter{page}{2340}

\maketitle

\label{firstpage}

\begin{abstract}
We examine the orbital dynamics of the planar pseudo-Newtonian Copenhagen problem, in the case of a binary system of Schwarzschild-like primaries, such as super-massive black holes. In particular, we investigate how the Jacobi constant (which is directly connected with the energy of the orbits) influences several aspects of the orbital dynamics, such as the final state of the orbits. We also determine how the relativistic effects (i.e., the Schwarzschild radius) affect the character of the orbits, by comparing our results with the classical Newtonian problem. Basin diagrams are deployed for presenting all the different basin types, using multiple types of planes with two dimensions. We demonstrate that both the Jacobi constant as well as the Schwarzschild radius highly influence the character of the orbits, as well as the degree of fractality of the dynamical system.
\end{abstract}


\section{Introduction}
\label{intro}

Black hole binary systems have become the subject of intense research since direct confirmation of their existence was provided by the recent discovery of the gravitational wave signals emitted from about a dozen black hole binary merger events by the LIGO collaboration \citep{AAA16a,AAA16b}. The black holes involved in these discoveries span a mass range going from $~10 M_\odot$ to $~100 M_\odot$, and are all consistent to have initially formed from the death of massive stars.

There is also strong observational support of the existence of a different population of so-called super-massive black holes (SMBHs), with masses spanning from $10^5$ to $10^{10}M_\odot$, residing at the center of virtually all galaxies \citep{BS12}. It is expected that some of these SMBHs will pair up as binaries following the merger of their host galaxies \citep{BBR80}. There is indeed evidence of several active galaxies with double nucleus \citep{KBH03,MSC15}, while the eventual inspiral and merger of some of these SMBH binaries constitutes a prime gravitational wave source for the planned LISA observatory \citep{ASA12}.

Under general conditions following a galaxy merger, the recently formed SMBH binary will typically be left with a binary separation of the order of 1 pc. This separation is at least two orders of magnitude larger than that needed for gravitational wave emission to be an efficient mechanism to drive the binary towards coalescence. This theoretical difficulty is known as the final parsec problem \citep{MM03}. It is considered that three-body interactions between the binary and individual stars in the circumnuclear cluster might play an important role for bridging this separation gap. Most of the studies that have been done of these systems have been based on non-relativistic analysis \citep{M02}. It is clear, however, that general relativistic effects should be dominant at least for a restricted region of the parameter space available to the interacting stars.

General relativity is a complex but successful geometric theory of gravitation that can be easily applied for the analytical modelling of some symmetric astrophysical scenarios of single compact objects \citep{SKM09}. However, for the simplest case of two massive bodies (i.e., SMBHs), there are no known exact solutions, only approximations or numerical solutions \citep[see e.g.,][]{PW14,BS10}. To avoid the high complexity of the equations of general relativity, analytical post-Newtonian or pseudo-Newtonian (PN) approaches are available \citep{UE99,SL06,B09}. The use of such approaches allows one to obtain approximate solutions and to reproduce some relativistic effects with some degree of accuracy.

Pseudo-Newtonian approaches have been widely used in astronomy and astrophysics \citep{M02}, motivated by the fact that its general predictions of particle dynamics do accurately agree with general relativistic computations \citep{CM06}. Since the seminal paper by Paczy\'{n}ski \& Wiita \citep{PW80}, various authors have proposed different PN potentials to model accretion disks around black holes \citep[see e.g.,][]{L98,ABN96,SK99}. In this context, it has been shown that these pioneer potentials have serious limitations when applied to non-circular trajectories, and hence, several alternative potentials have been developed recently that extend the range of validity of this kind of tools \citep[see e.g.,][]{WL17,TR13,W12}.

On the other hand, the study of the motion of test particles around binary black hole systems based on reduced three-body problem approaches (Roche approximations) is a relatively recent line of research \citep[see e.g.,][]{DLCG17}. In point of fact, in a recent paper, some of us have given a general theoretical framework based on the Paczy\'{n}ski \& Wiita potential for describing the gravitational field of two non-Newtonian primaries \citep{ZDG18}. The possibility of using PN potentials for studying the dynamics around binary black holes is justified by the fact that all PN approaches are just modifications to the Newtonian physics designed for mimicking some particular general-relativistic behaviours. In the present work, we examine the nature of motion of the planar PN Copenhagen problem, in the case of Schwarzschild-like primaries. The primaries are modelled with the aid of the Tejeda-Rosswog potential \citep{TR13}, which can reproduce exactly general trajectories of test particles in presence of a Schwarzschild black-hole. By adopting this PN potential we intend to capture and replicate some general relativistic effects, relevant to the Copenhagen problem. Moreover, we expect these effects to become more prominent and obvious, as we increase the value of the Schwarzschild radius. With the present work we expect to contribute to the understating of the role played by general relativistic effects in the dynamics of test particles around SMBH binaries. In particular, our numerical analysis has tangible astrophysical applications, since it describes the planar motion of a test particle in a binary system of two equally massed black holes, while it also demonstrates the influence of the general relativistic effects (i.e., the Schwarzschild radius) on the character of the orbits of the test particle

It should be stressed that for studying this particular problem, the following assumptions were made: i) The primaries move in circular closed loop orbits about their common mass center; ii) the total potential of the combined gravitational field can be expressed as the sum of the individual potentials of each primary; iii) the motion of the test particle takes place on the $(x,y)$ with $z = 0$; and iv) the binary separation is larger than the length scale over which the gravitational wave emission become relevant for the orbital dynamics.

The present article is, to a considerable extent, a continuation of the work initiated in \citet{ZDG18}, in which we determine the character of motion of a test particle, moving in a binary system of two compact objects. In Section \ref{des} the mathematical formulation of the astrophysical system, along with the disaggregation of the critical energy levels, are presented. In Section \ref{clas}, the basin diagrams are sketched on the configuration plane and then compared to the distribution of close encounter and escape times, spent by each orbit. Next, we discuss the parametric variation of the percentages of all types of orbits, classified in our study. We close this section by showing how the Schwarzschild radius affects the extent of the main types of basins: bounded, close encounter and escape. The most important conclusions of this numerical study are summarised in Section \ref{conc}.

\section{Pseudo-newtonian model}
\label{des}

Recently, in \cite{TR13} a new PN potential (TR herafter) has been introduced, which reproduces many relativistic features with better accuracy than many commonly used simple potentials, such as the Paczy\'{n}sky-Wiita (PW) \citep{PW80} and the Nowak-Wagoner (NW) \citep{NW91} potentials. In particular, the TR potential was derived in a self-consistent manner by taking the low-energy limit of the general relativistic equation for geodesic motion of test particles around a Schwarzschild black hole. This is in contrast to commonly used pseudo-Newtonian potentials (e.g., PW and NW) that are introduced in an ad-hoc basis in order to capture some particular relativistic feature. In particular, the PW potential was conceived with the express purpose of capturing the location of the marginally stable circular orbit (also known as innermost stable circular orbit or ISCO), but other relevant features are not so well reproduced (e.g., epicyclic frequencies and pericentre precession), not to mention that general particle motion (non-circular trajectories) cannot be reproduced with this potential. On the other hand, the TR potential captures several key relativistic effects exactly (e.g., the location of the marginally stable and marginally bound circular orbits, the radial dependence of the energy and angular momentum of circular orbits, the geometry of general orbits), and at the same time gives an accurate description of other relevant relativistic features (e.g., pericentre precession, epicyclic frequencies). More information on the performance of the TR potential as well as a comparison with other pseudo-Newtonian potentials can be found in \citet{TR13}.

It should also be noted that all three types of potentials (PW, NW and TR) coincide far away from the compact object, as they converge to the Newtonian potential, but in the relativistic regime their dynamical properties are totally different.

The TR potential, in spherical coordinates $(r,\theta,\phi)$, can be written as
\begin{equation}
\Phi(r,\theta,\phi) = \frac{G m}{r} + \left(\frac{2 r_g}{r_{S2}}\right)\left[\left(\frac{r_{S1}}{r_{S2}}\right)\dot{r}^2 + \frac{s(\theta,\dot{\theta},\dot{\phi})}{2}\right],
\label{pot0}
\end{equation}
where $r_g = G m/c^2$ corresponds to the gravitational radius, while
\begin{align}
&r_{S1} = r - r_g, \nonumber\\
&r_{S2} = r - 2 r_g, \nonumber\\
&s(r,\theta,\dot{\theta},\dot{\phi}) = r^2 \left(\dot{\theta}^2 + \dot{\phi}^2 \sin^2 \theta \right).
\end{align}
This potential contains two terms of which the first represents the usual, classical Newtonian potential, while the second term dictates the contributions due to the kinetic energy.

In this context, the two body problem can be formulated in the following way: it is assumed that both primaries have the same mass $(m_1 = m_2)$ and they freely move in a circle of radius $R/2$ around the barycenter, then we equate the radial force of gravity, exerted by each primary body on the other, with the centrifugal force, such that the angular velocity must satisfy
\begin{equation}
\omega = \sqrt{\frac{2G m \left(R - r_S\right)^2}{R^3 \left(R^2 - R r_S - r_S^2\right)}},
\label{ang}
\end{equation}
where $r_S = 2G m/c^2$ denotes the Schwarzschild radius. It should be noted, that by setting $r_S = 0$ Eq. (\ref{ang}) reduces to the classical form.

Using the standard notation for the normalization of constants in the planar circular restricted three-body problem, i.e., setting $G(m_1 + m_2) = 1$ for the masses and $x_1 + x_2 = 1$ for the distance between the primaries, the origin always coincide with the barycenter in the synodic system, such that introducing the parameter $\mu$,  we have that $G m_1 = 1 - \mu$, $G m_2 = \mu$, $x_1 = - \mu$ and $x_2 = 1 - \mu$. It should be noted, that the value $\mu = 1/2$ corresponds to the Copenhagen problem\footnote{For more details regarding the dimensionless units, we refer the reader to the reference \cite{S67}.}.

Hence, the PN effective potential that describes a relativistic binary system of two compact objects, in the equatorial plane, can be written in cartesian coordinates as
\begin{align}
&U(x,y,\dot{x},\dot{y}) = \frac{1 - \mu}{r_1} + \frac{\mu}{r_2} + \frac{\omega^2}{2} \left(x^2 + y^2\right) \nonumber\\
& + \frac{1}{2}r_S \bigg(\frac{\left(2r_1 - r_S\right)\left[\dot{x}\left(x - x_1\right) + y \dot{y}\right]}{r_1^2 \left(r_1 - r_S\right)^2} \nonumber\\
&+ \frac{\left(2r_2 - r_S\right)\left[\dot{x}\left(x - x_2\right) + y \dot{y}\right]}{r_2^2 \left(r_2 - r_S\right)^2} \nonumber\\
&+ \omega^2 \left(\frac{r_1^2}{r_1 - r_S} + \frac{r_2^2}{r_2 - r_S}\right) \bigg),
\label{pot}
\end{align}
with
\begin{align}
&r_1 = \sqrt{\left(x - x_1\right)^2 + y^2}, \nonumber\\
&r_2 = \sqrt{\left(x - x_2\right)^2 + y^2}.
\label{dist}
\end{align}

Now, we use Eq. (\ref{pot}) to construct the Lagrangian per unit mass in a synodic reference frame for the planar RTBP \citep[see e.g.,][]{S67}
\begin{equation}
\mathcal{L} = \frac{1}{2}\left(\dot{x}^2 + \dot{y}^2 \right) + \omega \left(x \dot{y} - \dot{x} y\right) + U(x,y,\dot{x},\dot{y}).
\label{lag}
\end{equation}

The respective equations of motion for the third body can be calculated with the aid of the Euler-Lagrange equations, as
\begin{align}
&\ddot{x} - 2 \omega \dot{y} = \frac{\partial U}{\partial x} - \frac{d}{dt}\left(\frac{\partial U}{\partial \dot{x}}\right), \nonumber\\
&\ddot{y} + 2 \omega \dot{x} = \frac{\partial U}{\partial y} - \frac{d}{dt}\left(\frac{\partial U}{\partial \dot{y}}\right).
\label{eqmot}
\end{align}

Due to the explicit dependence of the effective potential on the velocities, we get a system of two equations of the form
\begin{align}
&C_1(x,y) \ddot{x} + C_2(x,y) \ddot{y} + f\left(x,y,\dot{x},\dot{y}\right) = 0, \nonumber\\
&C_3(x,y) \ddot{x} + C_4(x,y) \ddot{y} + g\left(x,y,\dot{x},\dot{y}\right) = 0,
\label{sys1}
\end{align}
whose solutions are given by
\begin{align}
&\ddot{x} = \frac{g\left(x,y,\dot{x},\dot{y}\right) C_2(x,y) - f\left(x,y,\dot{x},\dot{y}\right) C_4(x,y)}{C_5(x,y)}, \nonumber\\
&\ddot{y} = \frac{f\left(x,y,\dot{x},\dot{y}\right) C_3(x,y) - g\left(x,y,\dot{x},\dot{y}\right) C_1(x,y)}{C_5(x,y)},
\label{sys2}
\end{align}
where $C_5 = C_1 C_4 - C_2 C_3$.

\begin{figure}
\centering
\resizebox{\hsize}{!}{\includegraphics{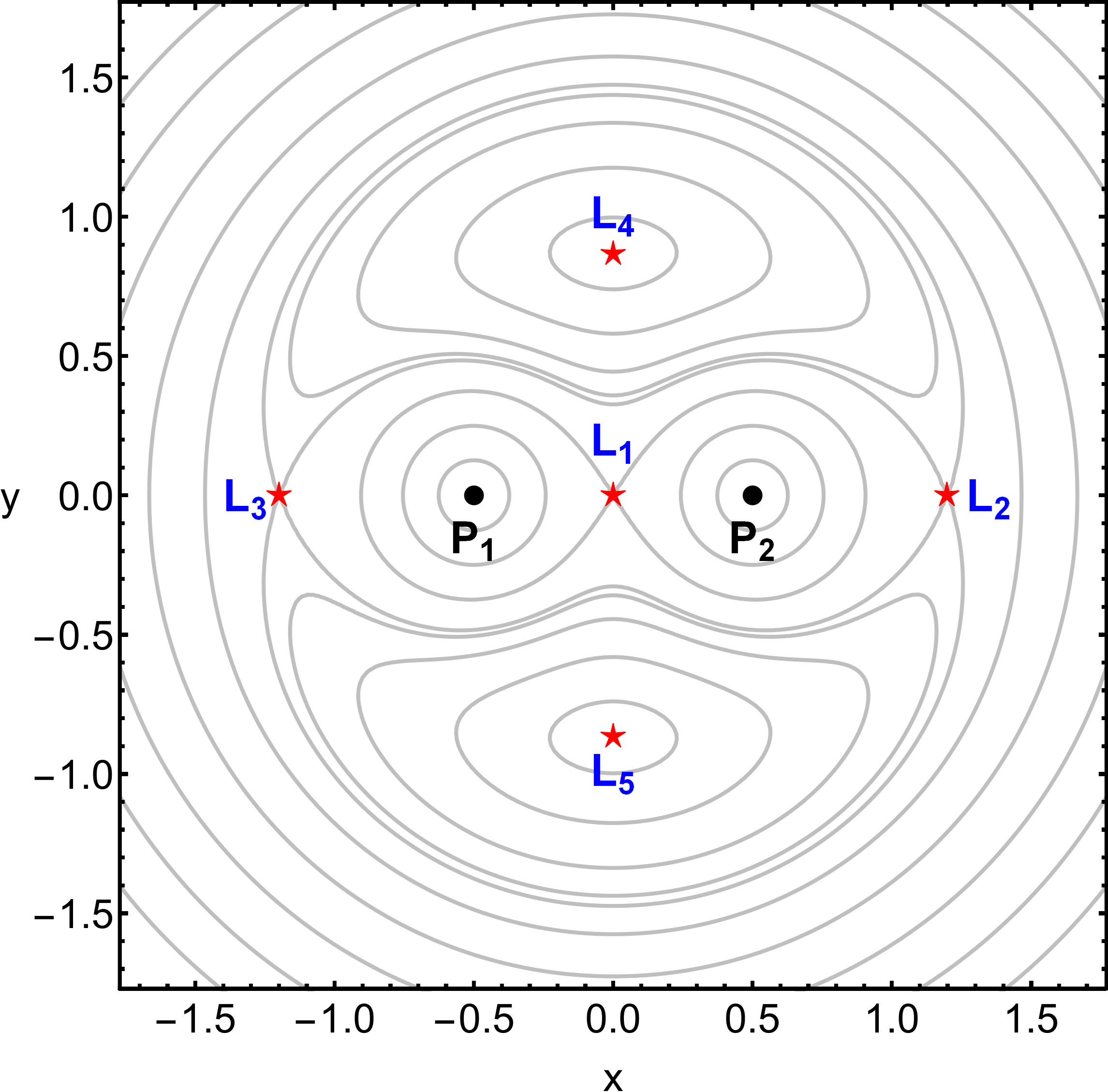}}
\caption{The contours of constant effective potential $U(x,y,\dot{x}=0,\dot{y}=0)$ on the plane $(x,y)$. Black dots are used for pinpointing the centres of the two compact objects, while red five-pointed stars mark the location of the five points of equilibrium of the system. (Color figure online.)}
\label{conts}
\end{figure}

The Jacobi integral is the only constant of motion and reads as
\begin{equation}
\mathcal{J} = 2U - \left(\dot{x}^2 + \dot{y}^2 \right) - 2 \left(\dot{x}\frac{\partial U}{\partial \dot{x}} + \dot{y}\frac{\partial U}{\partial \dot{y}} \right) = C.
\label{ham}
\end{equation}
Note, that unlike in the classical RTBP, the last term is not zero due to the velocity dependent terms in the PN effective potential (\ref{pot}).

In Appendix \ref{appex} we provide the explicit formulae for all the terms entering the equations of motion as well as the Jacobi integral.

The binary system of the two compact objects has five equilibrium points, all of them lying on the plane $(x,y)$. The firth three libration points, that is $L_i$, with $i = 1, ..., 3$, are collinear equilibrium points on the horizontal $x$-axis, while the other two $L_4$ and $L_5$ lie on the vertical $y$-axis (see Fig.~\ref{conts}). These equilibrium points divide the plane of motion $(x,y)$ into two areas: (i) the interior region (IR) for $x(L_3) < x < x(L_2)$ and the exterior region (ER) for $x < x(L_3)$ or $x > x(L_2)$.

The Jacobi values at the points of equilibrium are critical values and they are denoted as $C_i$, $i = 1,..., 4$. The geometry of the Hill's regions configuration on the plane $(x,y)$ is a function of these critical energy levels. More specifically:
\begin{itemize}
  \item Energy interval I $(C > C_1)$: The test particle can move either very close or very far from the primaries, while both IR and ER are disconnected.
  \item Energy interval II $(C_2 = C_3 < C < C_1)$: The third body  is allowed to circulate around the primary bodies in the IR, because the channel around the equilibrium point $L_1$ is open.
  \item Energy interval III $(C_4 = C_5 < C < C_2 = C_3)$: The test particle can move from IR to ER and vice versa, due to the open necks around the saddle points $L_2$ and $L_3$.
  \item Energy interval IV $(C < C_4 = C_5)$: The third body is allowed to move on the entire configuration space $(x,y)$, due to the complete absence of the energetically not allowed regions of motion.
\end{itemize}

\section{Classification of the initial conditions}
\label{clas}

\begin{figure*}
\centering
\resizebox{\hsize}{!}{\includegraphics{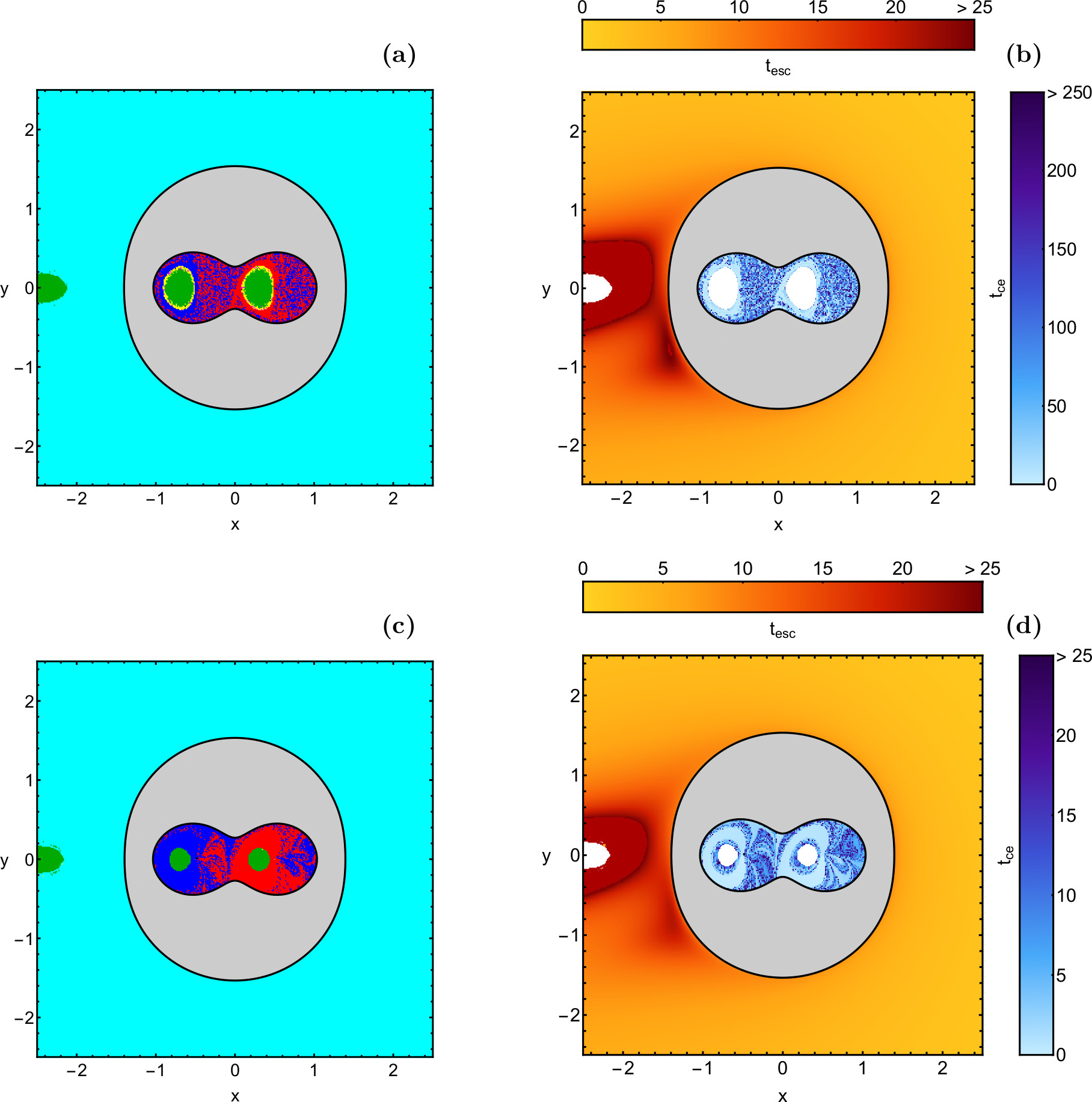}}
\caption{First column: The colour-coded basins of orbits on the $(x,y)$ plane, when $C = 3.6$. The colours denoting the different basin types are: ordered orbits (green); chaotic trapped orbits (yellow); close encounter motion to $P_1$ (blue); close encounter motion to $P_2$ (red); escaping motion (cyan); energetically not allowed regions (grey). Second column: The distribution of the respective close encounter and escape time of the orbits. First row: the classical problem with $r_S = 0$. Second row: the PN problem with $r_S = 0.01$. (Color figure online).}
\label{grd1}
\end{figure*}

\begin{figure*}
\centering
\resizebox{\hsize}{!}{\includegraphics{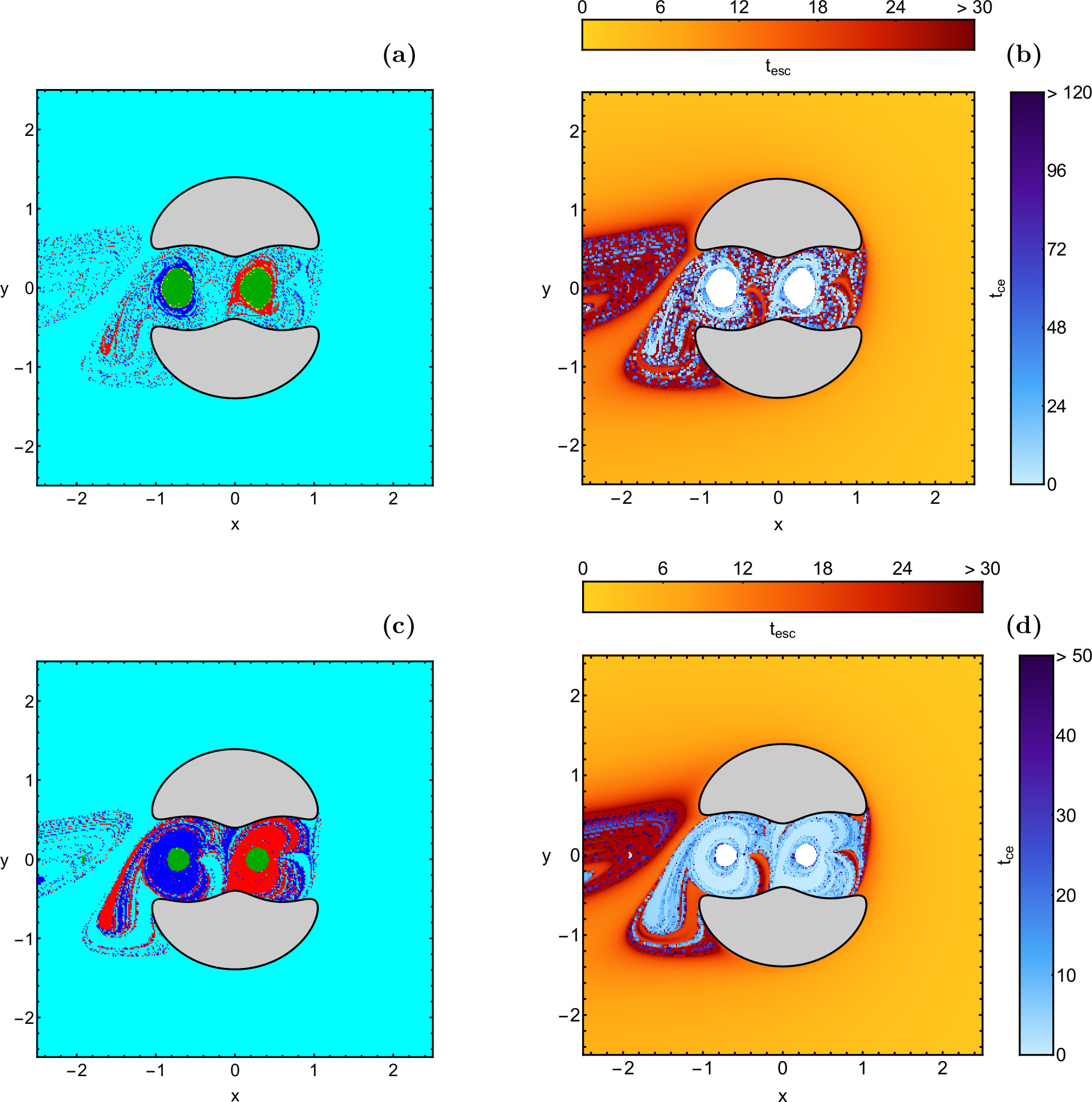}}
\caption{First column: The colour-coded basins of orbits on the $(x,y)$ plane, when $C = 3.3$. The colours denoting the different basin types are the same as in Fig.~\ref{grd1}. Second column: The distribution of the corresponding close encounter and escape time of the orbits. First row: the classical problem with $r_S = 0$. Second row: the PN problem with $r_S = 0.01$. (Color figure online).}
\label{grd2}
\end{figure*}

\begin{figure*}
\centering
\resizebox{\hsize}{!}{\includegraphics{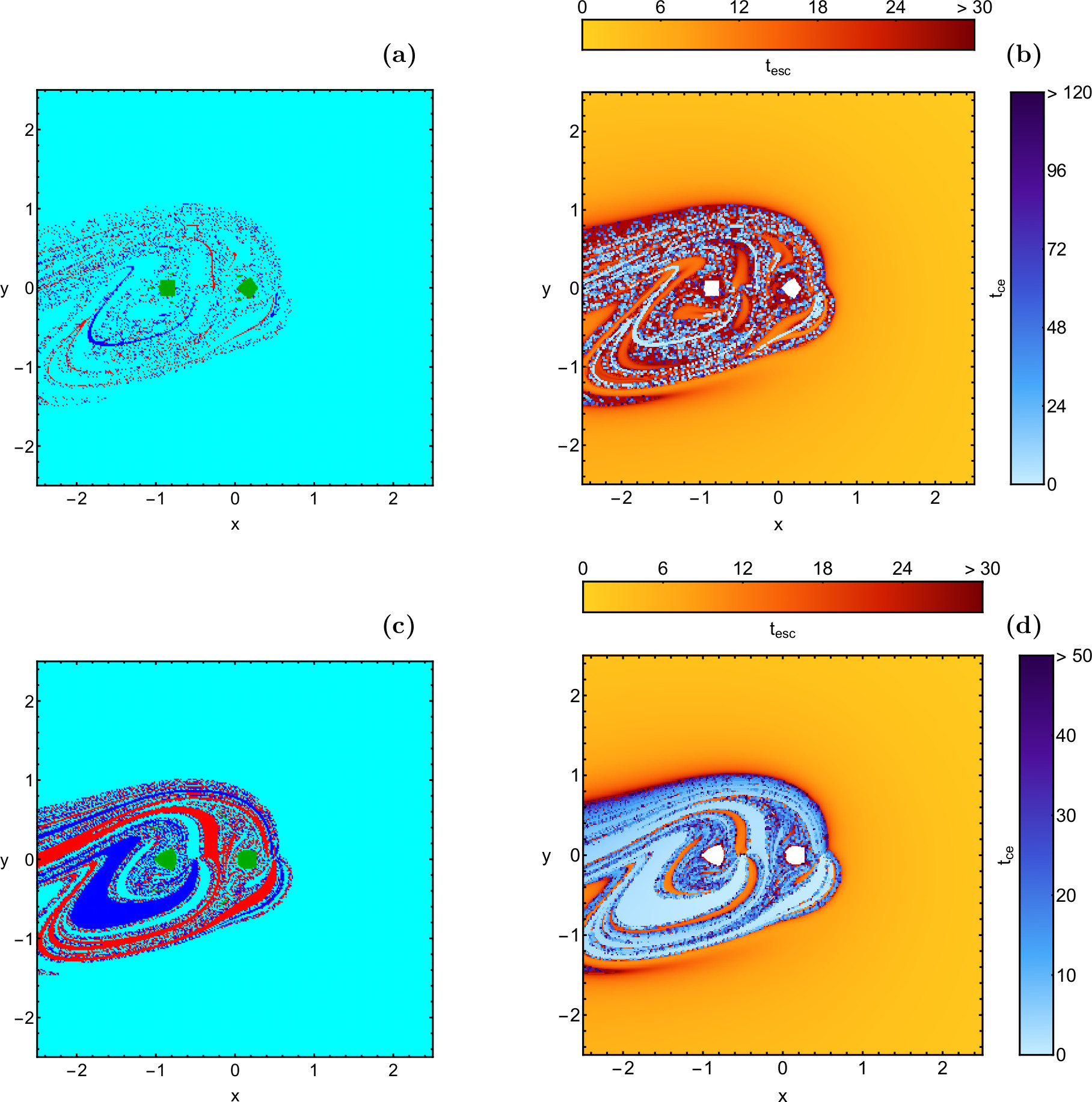}}
\caption{First column: The colour-coded basins of orbits on the $(x,y)$ plane, when $C = 2$. The colours denoting the different basin types are the same as in Fig.~\ref{grd1}. Second column: The distribution of the corresponding close encounter and escape time of the orbits. First row: the classical problem with $r_S = 0$. Second row: the PN problem with $r_S = 0.01$. (Color figure online).}
\label{grd3}
\end{figure*}

\begin{figure*}
\centering
\resizebox{\hsize}{!}{\includegraphics{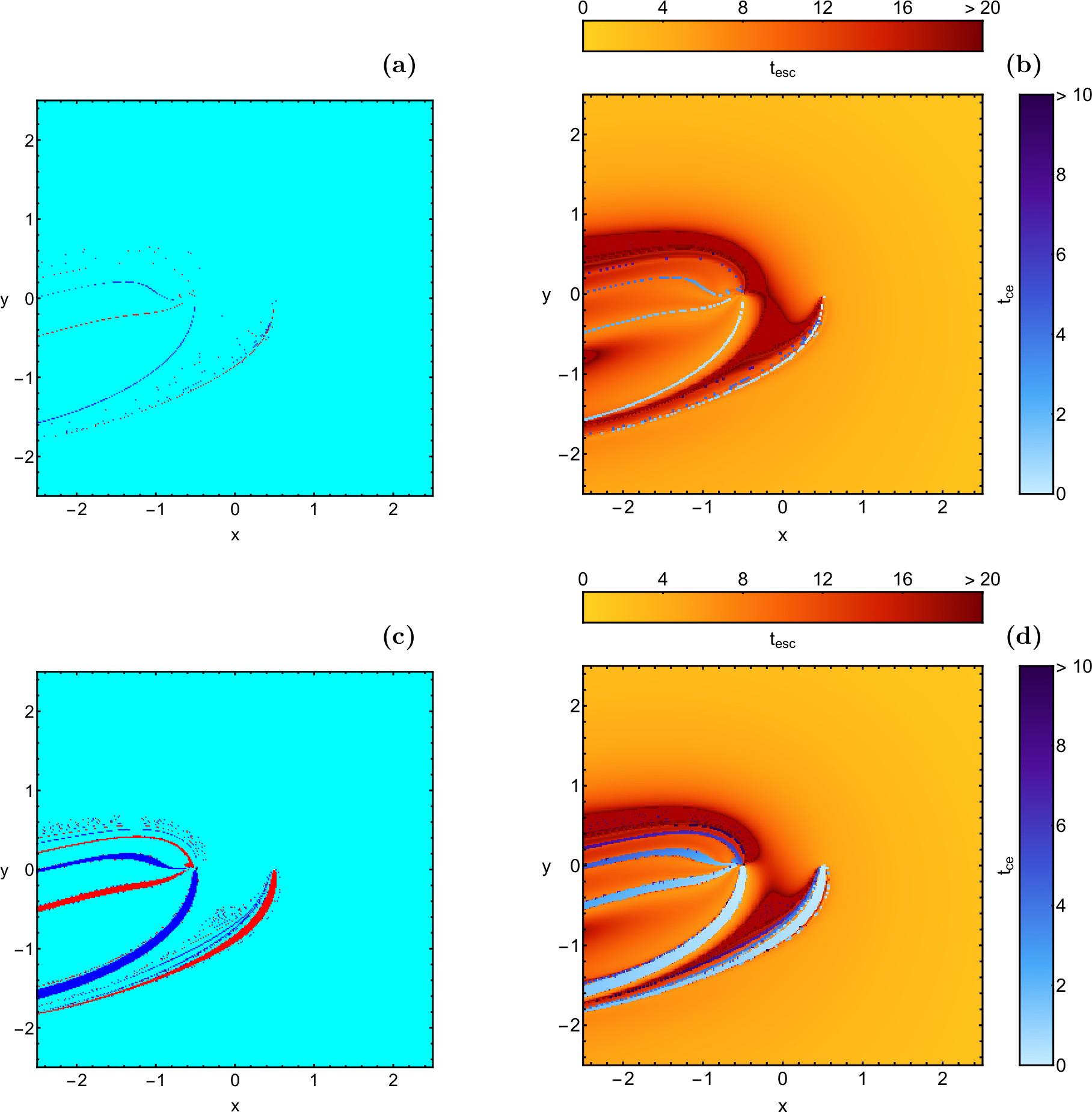}}
\caption{First column: The colour-coded basins of orbits on the $(x,y)$ plane, when $C = -0.5$. The colours denoting the different basin types are the same as in Fig.~\ref{grd1}. Second column: The distribution of the corresponding close encounter and escape time of the orbits. First row: the classical problem with $r_S = 0$. Second row: the PN problem with $r_S = 0.01$. (Color figure online).}
\label{grd4}
\end{figure*}

\begin{figure*}
\centering
\resizebox{\hsize}{!}{\includegraphics{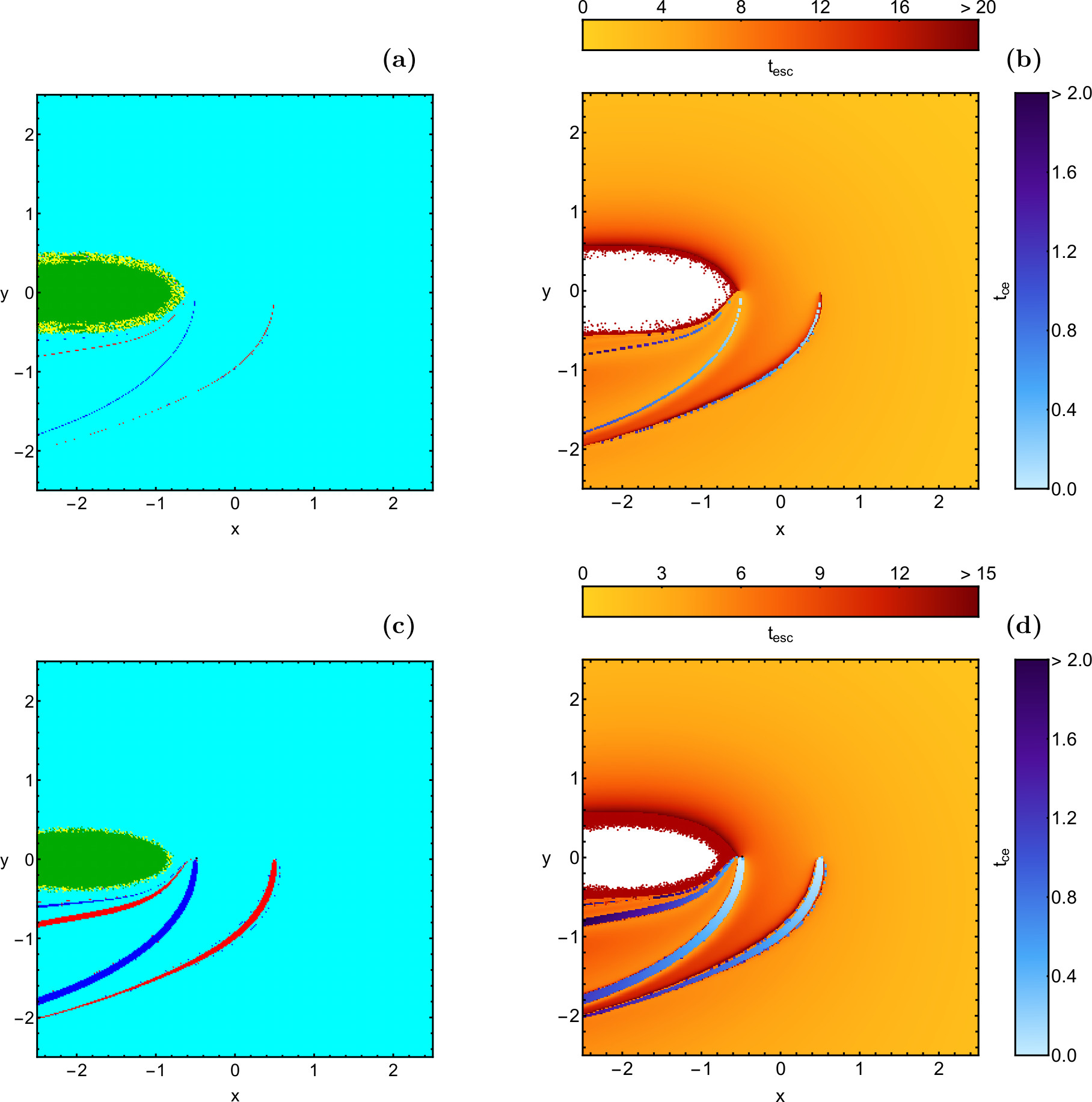}}
\caption{First column: The colour-coded basins of orbits on the $(x,y)$ plane, when $C = -2$. The colours denoting the different basin types are the same as in Fig.~\ref{grd1}. Second column: The distribution of the corresponding close encounter and escape time of the orbits. First row: the classical problem with $r_S = 0$. Second row: the PN problem with $r_S = 0.01$. (Color figure online).}
\label{grd5}
\end{figure*}

\begin{figure*}
\centering
\resizebox{\hsize}{!}{\includegraphics{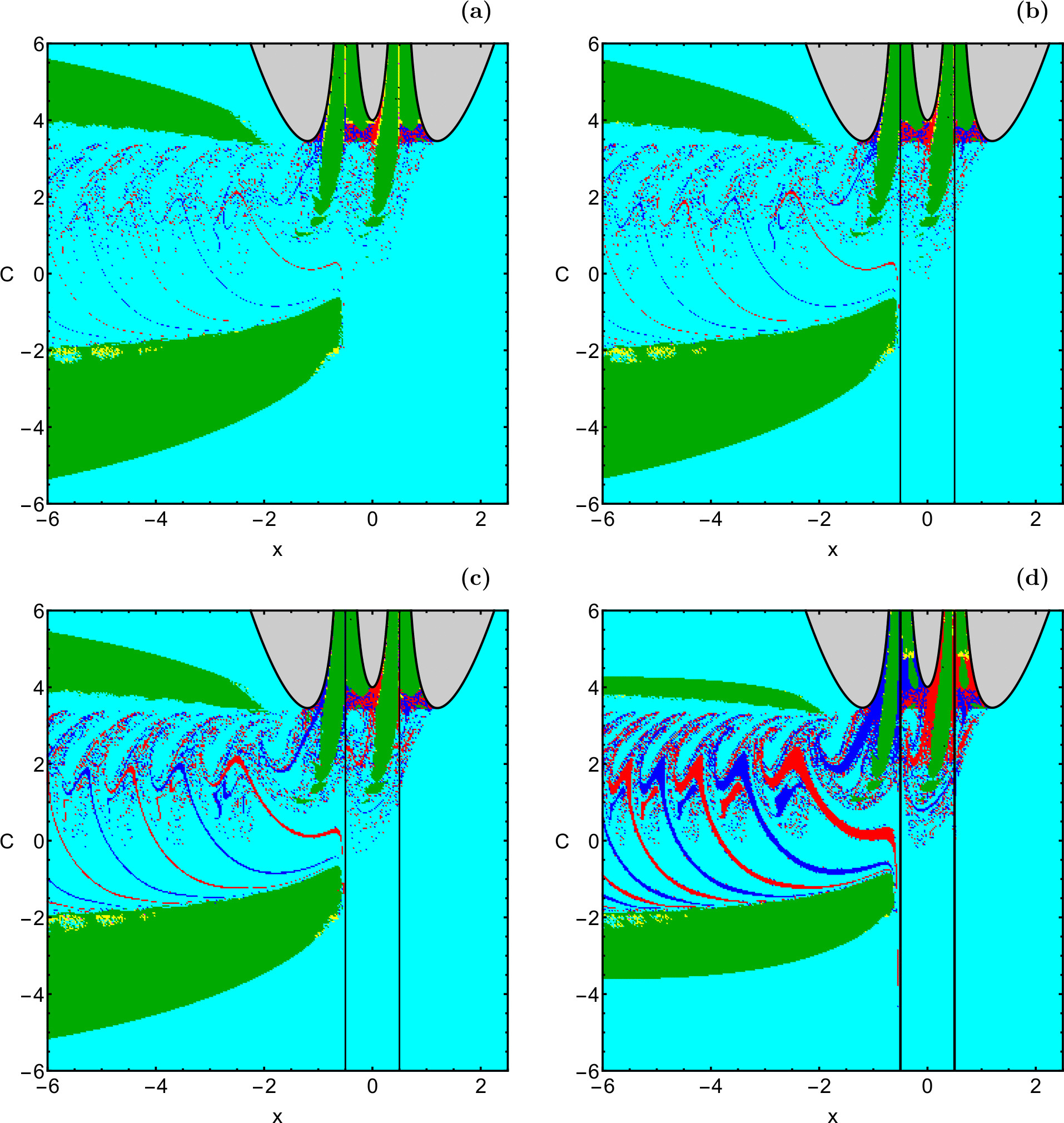}}
\caption{The colour-coded basins of orbits on the plane $(x,C)$, when (a-upper left): $r_S = 0$; (b-upper right): $r_S = 10^{-4}$; (c-lower left): $r_S = 10^{-3}$; (d-lower right): $r_S = 10^{-2}$. The colours denoting the different basin types are the same as in Fig.~\ref{grd1}. (Color figure online).}
\label{xC}
\end{figure*}

\begin{figure*}
\centering
\resizebox{\hsize}{!}{\includegraphics{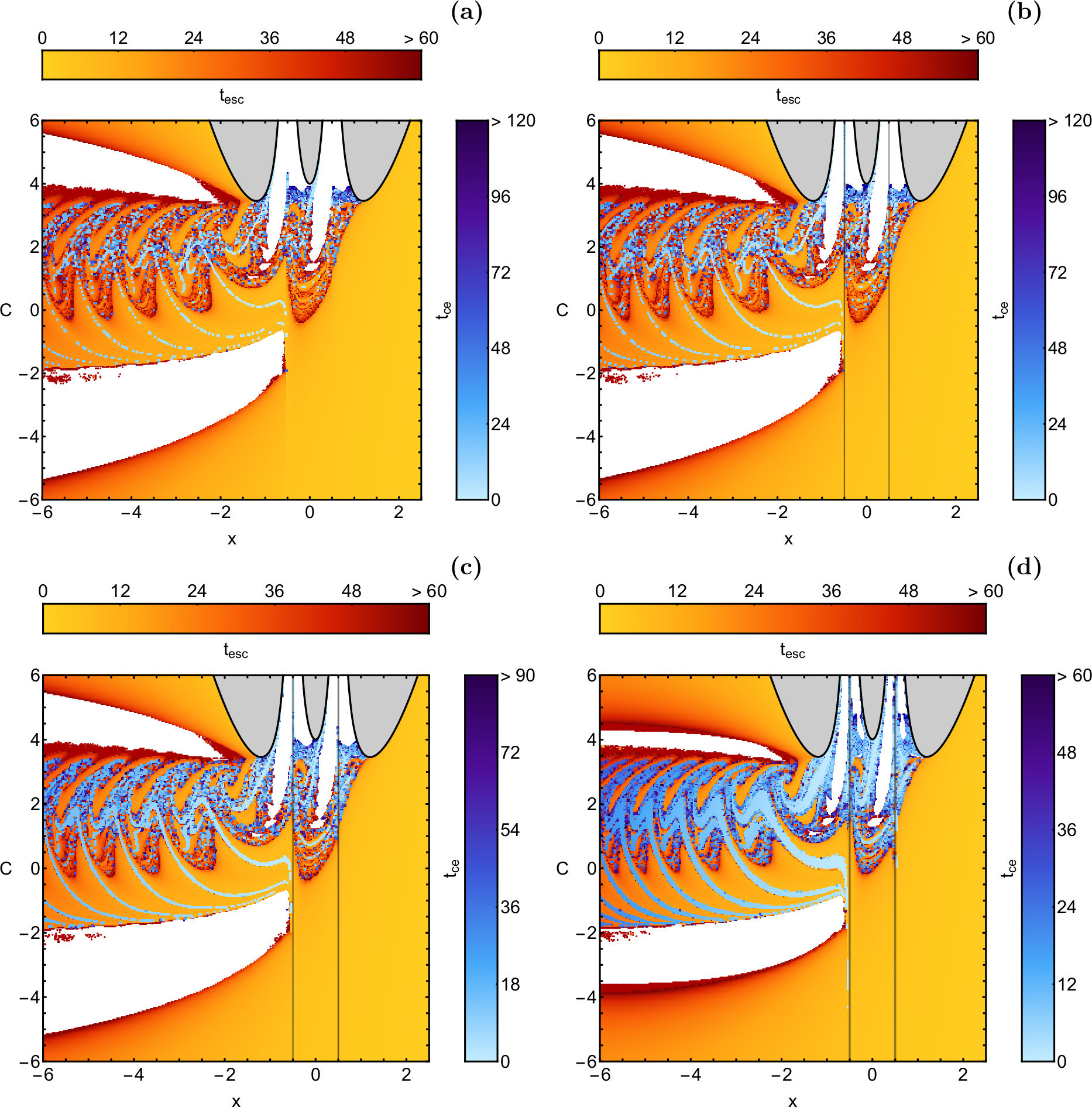}}
\caption{The respective distribution of the close encounter and escape time of the orbits, for the diagram shown in Fig.~\ref{xC}. (Color figure online).}
\label{xCt}
\end{figure*}

\begin{figure*}
\centering
\resizebox{\hsize}{!}{\includegraphics{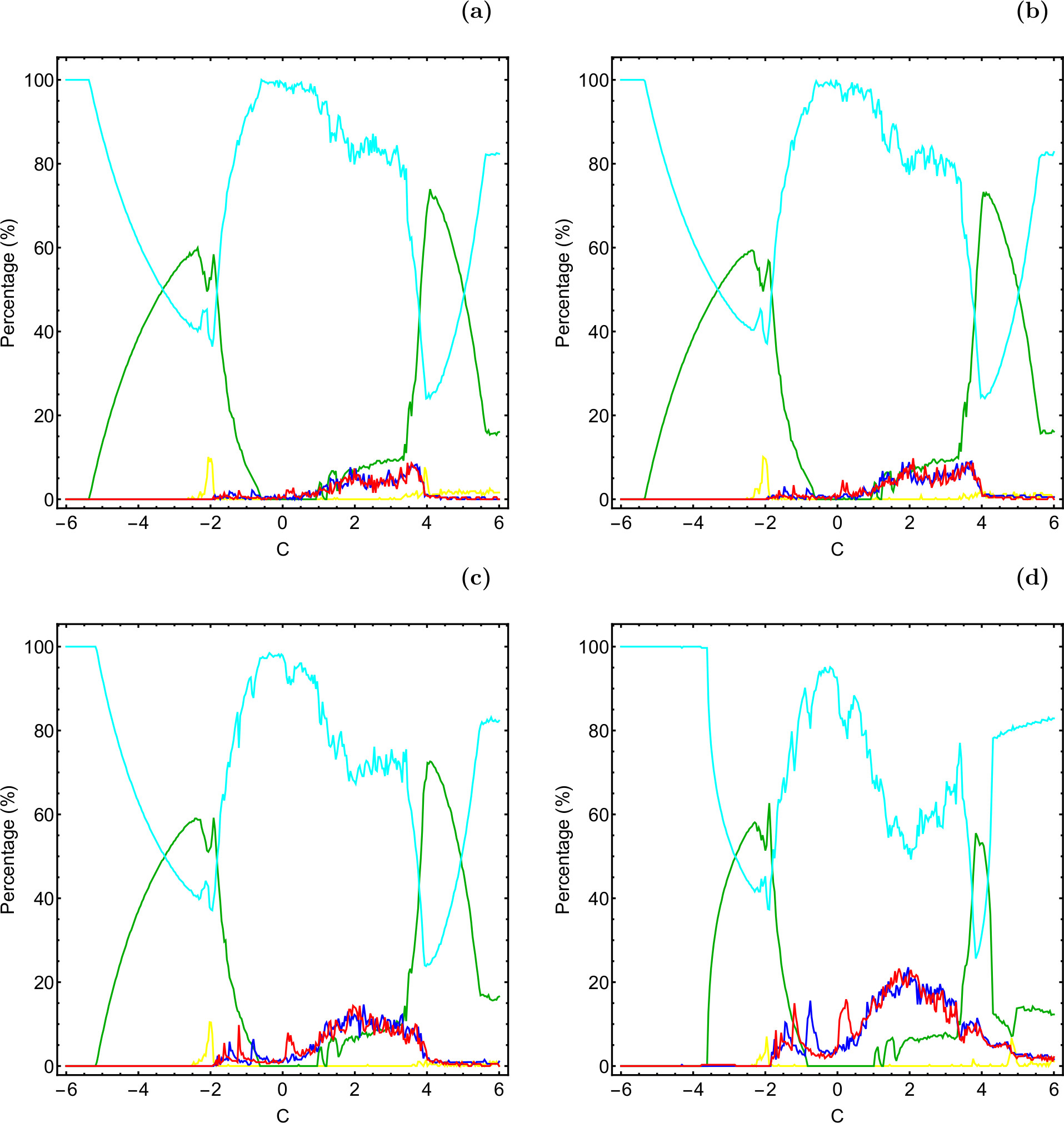}}
\caption{Evolution of the percentages of regular (green), escaping (cyan), and close encounter motion (blue, red), in terms of $C$, when (a): $r_S = 0$, (b): $r_S = 10^{-4}$, (c): $r_S = 10^{-3}$, and (d): $r_S = 10^{-2}$. (Color figure online).}
\label{pxC}
\end{figure*}

\begin{figure*}
\centering
\resizebox{\hsize}{!}{\includegraphics{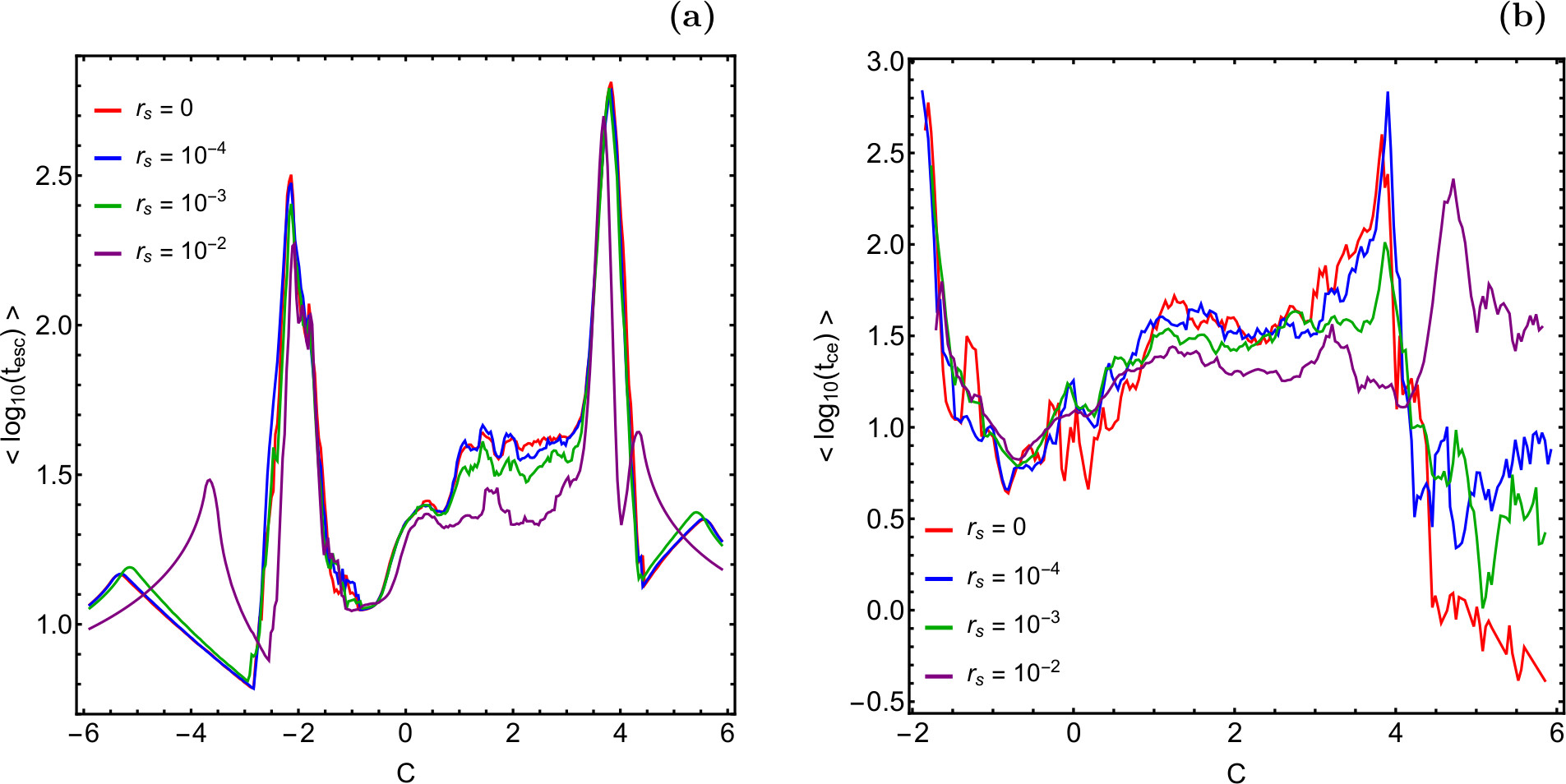}}
\caption{Semi-logarithmic representation of the average (a): escape and (b): close encounter times for initial conditions on the plane $(x,C)$, as a function of the Jacobi integral $C$. (Color figure online).}
\label{times}
\end{figure*}

\begin{figure}
\centering
\resizebox{\hsize}{!}{\includegraphics{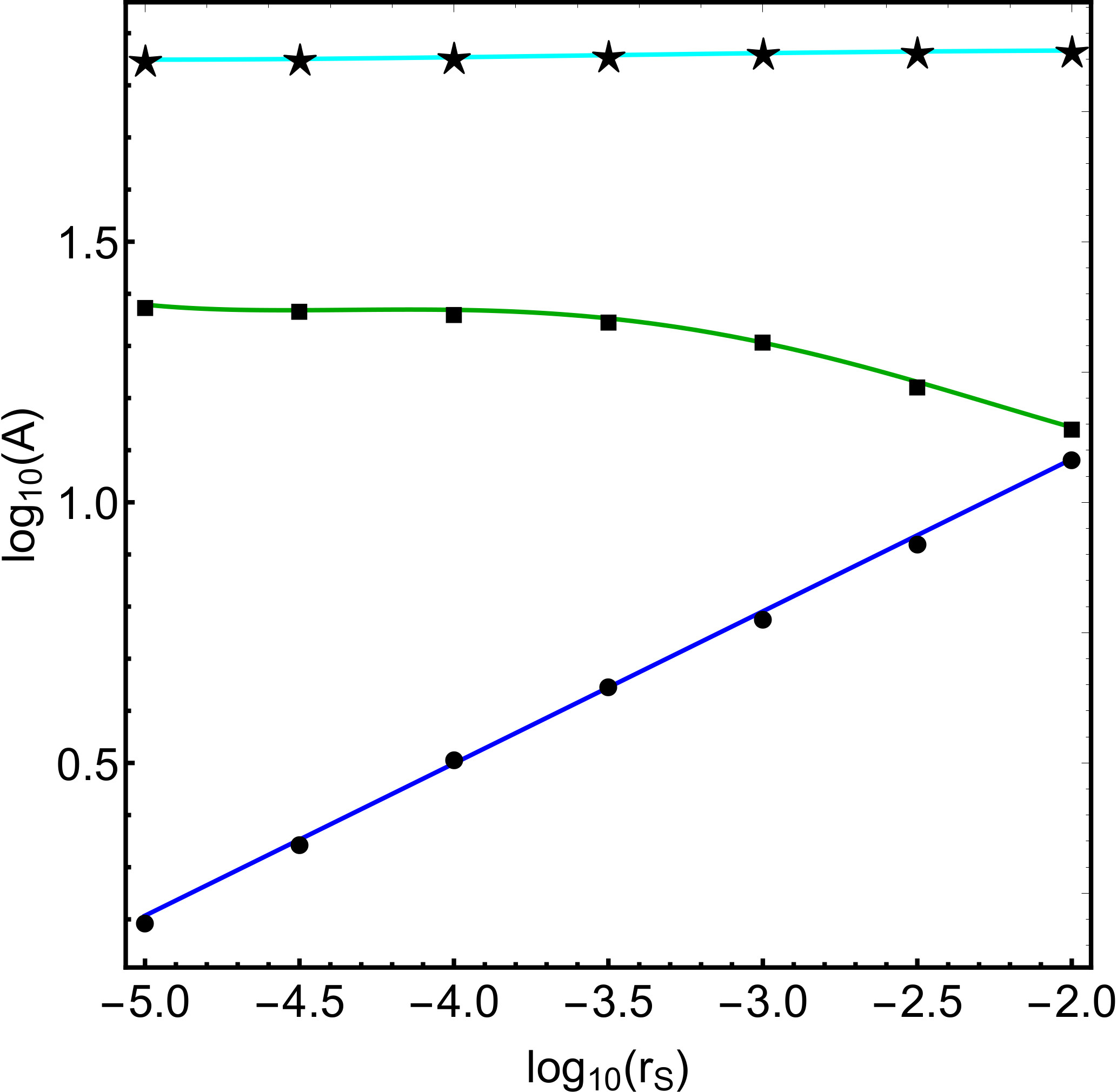}}
\caption{Diagram showing how the Schwarzschild radius $r_S$ of the primaries affects the extent of the main basin types (bounded, close encounter and escaping). The colours denoting the different types of orbits are: bounded regular orbits (green); close encounter orbits (blue); escaping orbits (cyan); energetically not allowed regions (grey).  (Color figure online).}
\label{rs}
\end{figure}

All orbits that we are going to classify start on the plane $(x,y)$ with starting velocities $\dot{x_0} = 0$, while in all cases $\dot{y_0}$ is derived through the Jacobi integral. The character of the orbits will be displayed by colouring the corresponding starting conditions $(x_0,y_0)$, according to the final state of the orbits. The computational methodology and all the numerical criteria we used for the taxonomy of the orbits are explained in detail in Section 4 of \citet{ZDG18}.

The following subsections correspond to the different Hill's regions configurations and of course to different Jacobi values. We chose to present the nature of motion of the configuration space $(x,y)$ for those Jacobi values, where the orbital content is interesting. It should be noted, that we do not present results for the case where $C > C_1$ (energy interval I), mainly for saving space, simply because the orbital content for this case is not interesting.

\subsection{Energy interval II $(C_2 = C_3 < C < C_1)$}
\label{ss1}

The first case corresponds to the second energy interval $C_2 = C_3 < C < C_1$, that is when the throat (channel) near the equilibrium point $L_1$ is open, which means that the test particle can move around both primary bodies, inside the IR. In Fig.~\ref{grd1} we present the orbit classification for $C = 3.6$, for the classical Copenhagen problem (panel (a)) and for the PN Copenhagen problem (panel (c)). In all the following similar figures the diagrams on the first row correspond to $r_S = 0$, while those on the second row correspond to $r_S = 0.01$. We chose to present the results for the extreme case with $r_S = 0.01$ so as to observe the highest impact of the PN dynamics on the nature of the orbits.

When $r_S = 0$ it is seen that inside the IR there exist two stability islands of regular bounded motion, which are formed by starting conditions of orbits moving around only one of the primary bodies. Just outside these bounded basins there are thin layers of chaotic trapped orbits, while the rest of the IR is filled with a highly chaotic mixture of close encounter orbits. One can argue, that relatively close to the primaries we can identify local basins of close encounter orbits. The ER is dominated by a unified area corresponding to escaping motion, while at the left side there is a small stability basin, composed of starting conditions of orbits circulating around $P_1$ and $P_2$.

For the PN problem, with $r_S \neq 0$, the orbital structure changes as follows:
\begin{itemize}
  \item The area of the bounded motion in both IR and ER is smaller.
  \item The chaotic layer, around the stability islands, disappears.
  \item The close encounter basins in the IR are more prominent.
\end{itemize}

Parts (b) and (d) of Fig.~\ref{grd1} show the way in which the times for close encounter and escape are distributed on the $(x,y)$ plane, using tones of red and blue, respectively. By inspecting the range of values at both colour bars it becomes evident that the time of escape of the orbits is almost unperturbed by the increase of $r_S$. On the other hand, the close encounter time of the orbits in the case of the PN problem is about ten times shorter, with respect to the classical problem. By the term time of escape $(t_{\rm esc})$ we refer to the time needed, so as the test particle to cross the escape radius $R_{\rm esc} = 10$ (with origin at $(0,0)$). In the same vein, the close encounter time $(t_{\rm ce})$ is the time required for a test particle to enter the disk with radius $R_c = r_S + r_{\rm ce}$, around each primary, with velocity pointing inwards. As in \citet{ZDG18}, $r_{\rm ce} = 10^{-5}$.

\subsection{Energy interval III $(C_4 = C_5 < C < C_2 = C_3)$}
\label{ss2}

The third energy interval $C_4 = C_5 < C < C_2 = C_3$ corresponds to the case where the two channels, in the vicinity of the points of equilibrium $L_2$ and $L_3$, are open and the test particle, coming from the IR, can enter the ER and vice versa. The character of motion on the $(x,y)$, for $C = 3.3$, is revealed in Fig.~\ref{grd2}. In the classical RTBP we see that the stability islands inside the IR are still there, while there no stability islands of regular motion around both primaries. With a closer look at panel (a) of Fig.~\ref{grd2} one can identify a small portion of trapped chaotic starting conditions. Furthermore, starting conditions of close encounter orbits exist also in the ER, forming thin trails, leading to the primaries.

When $r_S = 0.01$ we observe the following changes:
\begin{itemize}
  \item Regular bounded motion is only possible in the IR, however, the corresponding area of the bounded basins is smaller.
  \item There aren't signs of chaotic behaviour.
  \item Starting conditions which lead to close encounter to one of the primaries form well-defined basins, which extend beyond the IR.
\end{itemize}

Once more, according to parts (b) and (d) of Fig.~\ref{grd2}, far from the primaries, the diagrams of escape times are identical in both cases (classical and PN problems). On the opposite, the close encounter time spent by the orbits in the PN case is about 2.4 times shorter, compared to the close encounter time of the same starting conditions for the classical version of the problem.

\subsection{Energy interval IV $(C < C_4 = C_5)$}
\label{ss3}

The last energy interval corresponds to the case where the test particle can freely move to the plane $(x,y)$, without any restrictions of energetically not allowed regions of motion. From a dynamical point of view, this energy range is the most interesting region because the orbital content changes drastically, with respect to the energy level.

Fig.~\ref{grd3} shows the nature of orbits when $C = 2$. In both the classical and the PN cases the areas of regular bounded motion seem to be about the same. The only apparent difference is about the basins of the close encounter motion which in the case of the PN dynamics are much more defined. Also for this energy level the distribution of the time of escape of the orbits is exactly the same in both cases, while in the case of the PN dynamics the orbits need about 2.4 times less time to perform a close encounter to one of the primaries, with respect to the classical version of the problem.

The colour diagrams in Fig.~\ref{grd4} display the classification of the orbits when $C = -0.5$. For this energy range escaping motion is the dominant type of motion, while bounded motion around the primaries is no longer possible. Close encounters are allowed in both cases, however, in the PN problem the corresponding basins are larger, implying that in this case, the phenomenon of close encounters is more probable. As for the distribution of the close encounter and escape time of the orbits the diagrams shown in parts (b) and (d) of Fig.~\ref{grd4} indicate that in both cases the rates are quite similar.

When $C = -2$ we observe in Fig.~\ref{grd5} that the geometry of the plane $(x,y)$ changes again drastically. In both, the classical and the PN cases, stability islands corresponding to bounded regular motion around both primaries reappear, while at the boundaries of the bounded basins we observe the presence of starting conditions corresponding to chaotic orbits. It deserves mentioning, that when $r_S = 0.01$ the amount of trapped chaotic starting conditions is lower, with respect to the classical RTBP. Moreover, the close encounter basins are more prominent in the PN case. The diagrams in parts (b) and (b) of Fig.~\ref{grd5} with the distributions of the close encounter and escape time of orbits suggest that in both cases the rates are fairly similar, however in the case of the PN problem the orbits seem to escape a little bit faster with respect to the classical problem.

At this point, we would like to explain an issue regarding the geometry of the different basin types on the configuration space $(x,y)$. In \citet{ZDG18} the colour-coded diagrams had several symmetries (for example the area of the close encounter basins to primary $P_1$ was equal to the area of the close encounter basins to primary $P_2$), which were manifested by the use of polar coordinates. However, in the present study, the respective basins diagrams are completely asymmetric because in this case, we cannot use again polar coordinates. This is because the effective potential (\ref{pot}) is also a function of the velocities $\dot{x}$ and $\dot{y}$.

\subsection{An overview analysis}
\label{over}

Previously, we presented the nature of orbits on the $(x,y)$ plane for specific values of $C$. Now we want to obtain more information on how the Jacobi constant affects the character of motion. For this purpose, we present in Fig.~\ref{xC}(a-d) numerical outcomes on the $(x,C)$ space, for four characteristic values of $r_S$, when all orbits have $y_0 = \dot{x_0} = 0$, while $\dot{y_0}$ is always calculated through the Jacobi integral (\ref{ham}). The corresponding diagrams showing the distributions of the close encounter and escape time of the orbits are given in Fig.~\ref{xCt}(a-d).

As the value of $r_S$ increases the most notable changes on the $(x,C)$ plane are:
\begin{itemize}
  \item The areas of bounded regular and chaotic motion in both IR and ER are reduced.
  \item The extent of the close encounter basins increases.
  \item The fractal-like regions on the $(x,C)$ plane are reduced. More precisely, by computing the uncertainty dimension $D_0$ \citep[see e.g.,][]{AVS01,AVS09}, we found that the degree of fractality of the $(x,C)$ plane is indeed reduced. Our computations suggest that when $r_S = 0$ $D_0 = 1.4123$, when $r_S = 10^{-4}$ $D_0 = 1.4082$, when $r_S = 10^{-3}$ $D_0 = 1.3871$, while when $r_S = 10^{-2}$ $D_0 = 1.3107$.
  \item The time spent by the orbits in a close encounter seems to be reduced, while the escape time of the orbits seems to be unaffected by the shift of $r_S$.
\end{itemize}

From the diagrams shown in Fig.~\ref{xC} we can extract additional information, regarding the influence of the relativistic effects (i.e., the Schwarzschild radius) on the character of the orbits. In Fig.~\ref{pxC} we present, as a function of $C$, the parametric evolution of the rates of all types of orbits. It is seen, that the most interesting energy range is for $-2 < C < 4$. In particular, we observe that in this range with an increasing value of $r_S$ the percentage of orbits corresponding to close encounters increases, whereas the number of escaping orbits is reduced. Moreover, we see that for extremely low $(C < -5)$ or high $(C > 5)$ escaping motion completely dominates, despite of the particular value of $r_S$.

In the same vein, we present, as a function of $C$, in Fig.~\ref{times} the evolution of the average escape and close encounter times of the orbits, with starting conditions on the $(x,C)$ plane. Note that in both diagrams the scale of the times is logarithmic. One can see, that there are specific energy intervals where there is a clear hierarchy, regarding the influence of $r_S$. More specifically, for $-0.5 < C < 3.5$ $< \log_{10}\left(t_{\rm esc}\right) >$ is reduced, with increasing values of $r_S$. Similarly, for $1 < C < 4$ $< \log_{10}\left(t_{\rm ce}\right) >$ is also reduced with increasing value of $r_S$, while for $C > 4.2$ the trend is reversed. For the rest of the energy ranges the influence of $r_S$ is not very clear and we cannot deduce safe conclusions.

Our numerical analysis ends with Fig.~\ref{rs} in which we present how $r_S$ influences the areas of the main basin types (bounded, escape and close encounter). One can see, that in all cases the most extended basin corresponds to escaping motion, while the area of this type of basin is practically unperturbed by the shift of $r_S$. Moreover, the area of the bounded ordered orbits decreases for increasing value of $r_S$. On the other hand, for the close encounter motion, the corresponding numerical value of the area on the $(x,C)$ planes displays a linear increase, in a log-log scale. This means that we have a power law between the value of the area of close encounter orbits and the Schwarzschild radius. A similar behaviour (a power law) has also been observed in \citet{N04} and \citet{N05}, for the classical RTBP, and also in \citet{ZDG18}, for the Paczy\'{n}ski-Wiita potential.

In Fig.~\ref{rs} a power law $A_{\text{ce}} = r_S^{\gamma}$ (where $A_{\text{ce}}$ is the area corresponding to close encounter orbits), with $\gamma = 0.296$ is found. The power law behaviour can be explained by the following scaling arguments.

Using Kepler's problem we can roughly approximate the RTBP in the case where the third body is sufficiently close to one of the primaries, just before a close encounter occurs. In addition, all rotation terms can be neglected. Assuming $r_S \ll r_1 \ll r_2$, the problem reduces to Einstein's computation of Mercury's orbit, with an additional $1/r_1^3$ term that determines the precession rate of the Keplerian ellipse in the radial potential. For $r_1\ll 1$, we have $1/r_1^3 \gg 1/r_1$, such that the relativistic correction term dominates and according to \citet{E15} one can approximate the effective potential as
\begin{equation}
U \approx \frac{G L^2}{m c^2}\frac{1}{r_1^3},
\label{ein}
\end{equation}
with $L^2$ the square of the angular momentum, which is constant.

The energy $E$ of the (precessing) Keplerian ellipse is given by
\begin{equation}
E = - \frac{1}{r_a + r_p}\pm \left(\frac{2r_p r_a}{r_p + r_a}\right)^{1/2},
\label{kep1}
\end{equation}
where $r_p$ denotes the pericenter and $r_a$ the apocenter distance.

Following the arguments used in \citet{N04,N05}, we solve Eq. (\ref{kep1}) for $r_a$ that yields
\begin{equation}
r_a(r_p) = - r_p - \frac{E + r_p^2}{E^2 - 2r_p} - \frac{\left(r_p^4 + 2E r_p^2 + 2r_p\right)^{1/2}}{E^2 - 2r_p},
\label{kep2}
\end{equation}
where $E$ is the total energy.

In general, a close encounter occurs when the third body enters the circle of radius $R:=r_1$ around the singularity, $r_p \le R$. Hence, for $R\ll 1$, $A_{\text{ce}} \approx 2 \pi r_a(0) \left( r_a(0) - r_a(R)\right)$. Assuming that $r_p \ll 1$,  then Eq. (\ref{kep2}) displays, as a function of $r_p$, an approximated square root behaviour
\begin{equation}
r_a(r_p)\approx - \frac{1}{E} + \frac{(2 r_p)^{1/2}}{E^2}.
\label{approx3}
\end{equation}

Since we define close encounter through the condition $R = r_S$ the close encounter area $A_{\text{ce}}$, as a function of $r_s$, is given by
\begin{equation}
A_{\text{ce}} \sim r_S^{\beta},
\label{kep3}
\end{equation}
with the exponent $\beta = 1/2$. This approximation explains qualitatively the power law behavior, observed in Fig.~\ref{rs} but deviates to some extent from the numerical value of the exponent $\gamma = 0.296$.

\section{Concluding remarks}
\label{conc}

The scope of the article was a numerical investigation of the type of the orbits of the planar PN Copenhagen problem with Schwarzschild-like primaries, such as super-massive black holes. For describing the dynamical properties of the primaries we used the Tejeda-Rosswog (TR) potential, which reproduces many relativistic effects (e.g., the innermost stable circular orbit, non-circular trajectories, epicyclic frequencies and pericentre precession) with better accuracy, than many other simple potentials, such as the Paczy\'{n}ski \& Wiita (PW) and Nowak-Wagoner (NW). By integrating large sets of orbits, we obtained the several basin types on different versions of two-dimensional (2D) planes, by means of colour-coded basin diagrams. Moreover, we explored how the Jacobi constant $C$, as well as the relativistic effects (i.e., the Schwarzschild radius $r_S$) influence the nature of the orbits. It was demonstrated, that the both these quantities strongly affect the final states of the orbits, as well as the degree of fractality of the dynamical system.

The following list summarizes the main outcomes of our study:
\begin{enumerate}
  \item Regardless of the $r_S$ value, escaping orbits predominate for both extremely low and high values of $C$.
  \item In the PN Copenhagen problem, the basins corresponding to starting conditions of close encounter orbits are larger and better defined, while the amount of close encounter orbits grows, with increasing values of $r_S$.
  \item The relativistic effects (i.e., the Schwarzschild radius) also affects the bounded regular motion of the system. In particular, the area of the ordered basins corresponding to ordered orbits circulating around one or both primaries seems to reduce, as the value of $r_S$ increases.
  \item For $-0.5 < C < 3.5$, it was found that the time of escape of the orbits is reduced (at least by 15 time units) with increasing value of $r_S$. On the contrary, the close encounter time of the orbits either increases (for $C > 4.5$, at least by 25 time units) or decreases (for $1 < C < 4$, at leat by 30 time units) with increasing value of $r_S$, depending on the particular energy range.
  \item The fractal degree of the system is reduced with increasing value of $r_S$, since all the basin boundaries become smoother and sharper, while at the same time all the noisy (chaotic) regions are confined. The drop of the degree of fractality was also confirmed by the computation of the uncertainty dimension.
\end{enumerate}

In \citet{ZDG18} we conducted a similar numerical analysis classifying initial conditions of orbits, using the PW potential. On this basis, it would be very interesting to discuss the main similarities and differences of the PW and TR potentials. Here it should be noted, that the PW effective potential is only a function of the coordinates $x$ and $y$, so we were able to use a specialized choice of initial conditions, which express the symmetries of the system. The TR effective potential on the other hand (see Eq. (\ref{pot})), is also a function of the velocities $\dot{x}$ and $\dot{y}$, which means that we cannot use again the same type of initial conditions. Therefore, we cannot directly relate the numerical outcomes of the basin diagrams on the $(x,y)$ plane. Nevertheless, a comparison can be made by using the numerical information of the classification of orbits on the $(x,C)$ plane.

For the PW and TR potentials we have the following main similarities:
\begin{itemize}
  \item The stability islands of non-escaping regular motion appear at the same locations, i.e., between the two primaries and outside them.
  \item The area of the close encounter basins grow with increasing value of the Schwarzschild radius. Additionally, the increase of the close encounter area follow, for both potentials, a linear pattern (on a log-log scale).
  \item Both the average escape and the close encounter time of the orbits evolve very similarly (displaying similar patterns) for both types of the potentials, as a function of the Jacobi constant.
\end{itemize}

For the PW and TR potentials we have the following main differences:
\begin{itemize}
  \item In the case of the PW potential the area of the stability islands seems to be unaffected by the shift of the Schwarzschild radius. On the contrary, using the TR potential we observed that the area of the bounded basins decreases, with increasing value of the Schwarzschild radius.
  \item At the left hand side of the $(x,C)$ diagrams and between the stability basins the area of the close encounter basins is larger in the PW potential, with respect to the corresponding diagrams of the TR potential for all values of the Schwarzschild radius.
  \item For the PW potential, we found that for relatively high values of the Schwarzschild radius all three main basin types (bounded, escaping and close encounter) seem to converge to the same rate. In the case of the TR potential, the area of the bounded and close encounter basins do converge, while that of the escaping orbits remains almost unperturbed.
\end{itemize}

For numerically integrating the grids of starting conditions we used a Bulirsch-Stoer routine in the standard version of \verb!FORTRAN 77! \citep[e.g.,][]{PTVF92}, with double precision. In all calculations the error, regarding the conservation of the Jacobi constant, was of the order of $10^{-13}$, or even smaller. Using an Intel Quad-Core i7 vPro 4.0 GHz processor, we needed between 3 and 11 hours of CPU time, for the taxonomy of orbits, per grid. All the graphics of the article have been constructed by using the 11.3 version of the Mathematica$^{\circledR}$ software \citep{W03}.

\section*{Acknowledgments}

The authors would like to express their warmest thanks to the anonymous referee for all the apt suggestions and comments which improved both the quality as well as the clarity of the paper.

\appendix

\section{Detailed presentation of the involved equations}
\label{appex}

Below we present the analytic expressions of all the coefficients, entering the equations of motion:
\begin{align}
&C_1(x,y) = 1 + r_S \bigg[ \frac{\left(2r_1 - r_S\right)\left(x + \mu\right)^2}{r_1^2 \left(r_1 - r_S\right)^2} \nonumber\\
&+ \frac{\left(2r_2 - r_S\right)\left(x + \mu - 1\right)^2}{r_2^2 \left(r_2 - r_S\right)^2},
\label{c1}
\end{align}
\begin{align}
&C_2(x,y) = C_3(x,y) = r_S y \bigg[ \frac{\left(2r_1 - r_S\right)\left(x + \mu\right)}{r_1^2 \left(r_1 - r_S\right)^2} \nonumber\\
&+ \frac{\left(2r_2 - r_S\right)\left(x + \mu - 1\right)}{r_2^2 \left(r_2 - r_S\right)^2} \bigg],
\label{c2}
\end{align}
\begin{equation}
C_4(x,y) = 1 + r_S y \left[ \frac{2r_1 - r_S}{r_1^2\left(r_1 - r_S\right)^2} + \frac{2r_2 - r_S}{r_2^2\left(r_2 - r_S\right)^2} \right],
\label{c4}
\end{equation}
\begin{align}
&C_5(x,y) = 1 + r_S \bigg[\frac{2r_1 - r_S}{\left(r_1 - r_S\right)^2} + \frac{2r_2 - r_S}{\left(r_2 - r_S\right)^2} \nonumber\\
&+ \frac{r_S y^2 \left(2r_1 - r_S\right)\left(2r_2 - r_S\right)}{r_1^2 r_2^2 \left(r_1 - r_S\right)^2 \left(r_2 - r_S\right)^2} \bigg],
\label{c5}
\end{align}
while the velocity dependent functions $f$ and $g$ are given by
\begin{align}
&f(x,y,\dot{x},\dot{y}) = \frac{\left(1 - \mu\right)\left(x + \mu\right)}{r_1^3} + \frac{\mu \left(x + \mu - 1 \right)}{r_2^3} - x \omega^2 - 2 \omega \dot{y} \nonumber\\
&+ r_S \bigg[ \frac{1}{2}\omega \left(- \frac{\left(r_1 - 2r_S\right)\left(x + \mu\right)}{\left(r_1 - r_S\right)^2} - \frac{\left(r_2 - 2r_S\right)\left(x + \mu - 1\right)}{\left(r_2 - r_S\right)^2}\right) \nonumber\\
&+ \frac{x + \mu}{r_1^4 \left(r_1 - r_S\right)^3} \bigg[r_1^2 \left(r_1 - r_S\right)\left(2r_1 - r_S\right)\left(\dot{x}^2 + \dot{y}^2\right) \nonumber\\
&- \left(3r_1^2 - 3r_1 r_S + r_S^2\right)\left(\dot{x}\left(x + \mu\right) + y\dot{y}\right)^2 \bigg] \nonumber\\
&+ \frac{x + \mu - 1}{r_2^4 \left(r_2 - r_S\right)^3} \bigg[r_2^2 \left(r_2 - r_S\right)\left(2r_2 - r_S\right)\left(\dot{x}^2 + \dot{y}^2\right) \nonumber\\
&- \left(3r_2^2 - 3r_2 r_S + r_S^2\right)\left(\dot{x}\left(x + \mu - 1\right) + y\dot{y}\right)^2 \bigg] \bigg],
\label{ff}
\end{align}
and
\begin{align}
&g(x,y,\dot{x},\dot{y}) = y \left(\frac{1 - \mu}{r_1^3} + \frac{\mu}{r_2^3}\right) - y \omega^2 + 2 \omega \dot{x} \nonumber\\
&+r_S y \bigg[ - \frac{1}{2}\omega^2 \left(\frac{r_1 - 2 r_S}{\left(r_1 - r_S\right)^2} + \frac{r_2 - 2r_S}{\left(r_2 - r_S\right)^2}\right) \nonumber\\
&+ \frac{1}{r_1^4\left(r_1 - r_S\right)^3} \bigg[ r_1^2 \left(r_1 - r_S\right)\left(2 r_1 - r_S\right)\left(\dot{x}^2 + \dot{y}\right) \nonumber\\
&- \left(3r_1^2 - 3r_1 r_S + r_S^2\right)\left(\dot{x}\left(x + \mu\right)+ y\dot{y}\right)^2 \bigg] \nonumber\\
&+ \frac{1}{r_2^4\left(r_2 - r_S\right)^3} \bigg[ r_2^2 \left(r_2 - r_S\right)\left(2 r_2 - r_S\right)\left(\dot{x}^2 + \dot{y}\right) \nonumber\\
&- \left(3r_2^2 - 3r_2 r_S + r_S^2\right)\left(\dot{x}\left(x + \mu - 1\right)+ y\dot{y}\right)^2 \bigg] \bigg].
\label{gg}
\end{align}

The explicit form of the Jacobi integral of motion reads
\begin{align}
&\mathcal{J} = 2\left(\frac{1 - \mu}{r_1} + \frac{\mu}{r_2}\right) + \omega^2 \left(x^2 + y^2\right) - \left(\dot{x}^2 + \dot{y}^2\right) \nonumber\\
&+ r_S \bigg[ \omega^2 \left(\frac{r_1^2}{r_1 - r_S} + \frac{r_2^2}{r_2 - r_S}\right) \nonumber\\
&- \frac{\left(2r_1 - r_S\right)\left(y\dot{y} + \dot{x}\left(x + \mu\right)\right)^2}{r_1^2\left(r_1 - r_S\right)^2} \nonumber\\
&- \frac{\left(2r_2 - r_S\right)\left(y\dot{y} + \dot{x}\left(x + \mu - 1\right)\right)^2}{r_2^2\left(r_2 - r_S\right)^2} \bigg].
\label{jac}
\end{align}
As can be noted from the previous expressions, in the limit $r_S \to 0$, the equations of motion (\ref{eqmot}) and the Jacobian (\ref{ham}) both reduce to their Newtonian counterparts.

\bsp
\label{lastpage}


\begin{thebibliography}{}

\bibitem[\protect\citeauthoryear{Abbott et al.}{2016a}]{AAA16a} Abbott B.P., Abbott R., Abbott T.D., Abernathy M.R., Acernese F., Ackley K., Adams C., Adams T., Addesso P., Adhikari R.X., et al., 2016a, Physical Review Letters, 116, 061102

\bibitem[\protect\citeauthoryear{Abbott et al.}{2016b}]{AAA16b} Abbott B.P., Abbott R., Abbott T.D., Abernathy M.R., Acernese F., Ackley K., Adams C., Adams T., Addesso P., Adhikari R.X., et al., 2016b, Physical Review X, 6, 041015

\bibitem[\protect\citeauthoryear{Aguirre et al.}{2001}]{AVS01} Aguirre J., Vallego J.C., Sanju\'{a}n M.A.F., 2001, Phys. Rev. E, 64, 066208

\bibitem[\protect\citeauthoryear{Aguirre et al.}{2009}]{AVS09} Aguirre J., Viana R.L., Sanju\'{a}n M.A.F., 2009, Rev. Mod. Phys., 81, 333

\bibitem[\protect\citeauthoryear{Amaro-Seoane et al.}{2012}]{ASA12} Amaro-Seoane P., Aoudia S., Babak S., Bin\'{e}truy P., Berti E., Boh\'{e} A., Caprini C., Colpi M., Cornish N. J., Danzmann K., et al. 2012, Classical and Quantum Gravity, 29, 124016

\bibitem[\protect\citeauthoryear{Artemova et al.}{1996}]{ABN96} Artemova I.V., Bjoernsson G., Novikov I.D., 1996, ApJ, 461, 565

\bibitem[\protect\citeauthoryear{Baumgarte \& Shapiro}{2010}]{BS10} Baumgarte T.W., Shapiro S.L., 2010, Numerical relativity: solving Einstein's equations on the computer. Cambridge University Press

\bibitem[\protect\citeauthoryear{Beckmann \& Shrader}{2010}]{BS12} Beckmann V., Shrader C. R., 2012, Active Galactic Nuclei, Wiley-VCH Verlag GmbH

\bibitem[\protect\citeauthoryear{Begelman et al.}{2010}]{BBR80} Begelman M.C., Blandford R.D., Rees M.J., 1980, Nature, 287, 307

\bibitem[\protect\citeauthoryear{Blanchet}{2009}]{B09} Blanchet L., 2009, Post-Newtonian theory and the two-body problem. In Mass and Motion in General Relativity (pp. 125-166). Springer, Dordrecht

\bibitem[\protect\citeauthoryear{Chakrabarti \& Mondal}{2006}]{CM06} Chakrabarti S.K., Mondal, S., 2006, MNRAS, 369, 976

\bibitem[\protect\citeauthoryear{Dubeibe et al.}{2017}]{DLCG17} Dubeibe F.L., Lora-Clavijo F.D., Gonz\'{a}lez G.A., 2017, Physics Letters A, 381, 563

\bibitem[\protect\citeauthoryear{Einstein}{1915}]{E15} Einstein A., 1915, ``Erkl{\"a}rung der Perihelbewegung des Merkur aus der allgemeinen Relativit{\"a}tstheorie", Sitzungsberichte der Preussischen Akademie der Wissenschaften, 47, 803.

\bibitem[\protect\citeauthoryear{Komossa et al.}{2003}]{KBH03} Komossa S., Burwitz V., Hasinger G., Predehl P., Kaastra J.S., Ikebe Y., 2003, Astrophysical Journal Letters, 582, L15

\bibitem[\protect\citeauthoryear{L{\o}v{\aa}s}{1998}]{L98} L{\o}v{\aa}s T., 1998, International Journal of Modern Physics D, 7, 471

\bibitem[\protect\citeauthoryear{Merritt \& Poon}{2002}]{MP04} Merritt D., Poon M.Y., 2004, Astrophysical Journal, 606, 788

\bibitem[\protect\citeauthoryear{Milosavljevi\'{c} \& Merritt}{2002}]{MM03} Milosavljevi\'{c} M., Merritt D., 2003, in Centrella J.M., ed., The Astrophysics of Gravitational Wave Sources Vol. 686 of American Institute of Physics Conference Series, The Final Parsec Problem. pp 201-210

\bibitem[\protect\citeauthoryear{Mukhopadhyay}{2002}]{M02} Mukhopadhyay B., 2002, ApJ, 581, 427

\bibitem[\protect\citeauthoryear{M\"{u}ller et al.}{2015}]{MSC15} M\"{u}ller-S\'{a}nchez F., Comerford J.M., Nevin R., Barrows R.S., Cooper M.C., Greene J.E., 2015, Astrophysical Journal, 813, 103

\bibitem[\protect\citeauthoryear{Nagler}{2004}]{N04} Nagler J., 2004, Phys. Rev. E, 69, 066218

\bibitem[\protect\citeauthoryear{Nagler}{2005}]{N05} Nagler J., 2005, Phys. Rev. E, 71, 026227

\bibitem[\protect\citeauthoryear{Nowask \& Wagoner}{1991}]{NW91} Nowak M. A., Wagoner R.V., 1991, ApJ, 378, 656

\bibitem[\protect\citeauthoryear{Paczy\'{n}ski \& Wiita}{1980}]{PW80} Paczy\'{n}ski, B., Wiita, P.J., 1980, A\&A, 88, 23

\bibitem[\protect\citeauthoryear{Poisson \& Will}{2014}]{PW14} Poisson E., Will C.M., 2014 Gravity: Newtonian, Post-Newtonian, Relativistic. Cambridge University Press

\bibitem[\protect\citeauthoryear{Press}{1992}]{PTVF92} Press H.P., Teukolsky S.A, Vetterling W.T., Flannery B.P., 1992, Numerical Recipes in FORTRAN 77, 2nd Ed., Cambridge Univ. Press, Cambridge, USA

\bibitem[\protect\citeauthoryear{Semerak \& Karas}{1999}]{SK99} Semerak O., Karas V., 1999, A\&A, 343, 325

\bibitem[\protect\citeauthoryear{Steklain \& Letelier}{2006}]{SL06} Steklain A.F., Letelier P.S., 2006, Physics Letters A, 352, 398

\bibitem[\protect\citeauthoryear{Stephani et al.}{2009}]{SKM09} Stephani H., Kramer D., MacCallum M., Hoenselaers C., Herlt E., 2009, Exact solutions of Einstein's field equations. Cambridge University Press

\bibitem[\protect\citeauthoryear{Szebehely}{1967}]{S67} Szebehely V., 1967, Theory of Orbit: The Restricted Problem of Three Bodies. Academic Press, New York

\bibitem[\protect\citeauthoryear{Tejeda \& Rosswog}{2013}]{TR13} Tejeda E., Rosswog S., 2013, MNRAS, 433, 1930

\bibitem[\protect\citeauthoryear{Ury\~{u} \& Eriguchi}{1999}]{UE99} Ury\~{u} K., Eriguchi Y., 1999, MNRAS, 303, 329

\bibitem[\protect\citeauthoryear{Wegg}{2012}]{W12} Wegg C., 2012, ApJ, 749, 183

\bibitem[\protect\citeauthoryear{Witzany \& L\"{a}mmerzahl}{2017}]{WL17} Witzany V., L\"{a}mmerzahl C., 2017, ApJ, 841, 105

\bibitem[\protect\citeauthoryear{Wolfram}{2003}]{W03} Wolfram S., 2003, The Mathematica Book. Wolfram Media, Champaign

\bibitem[\protect\citeauthoryear{Zotos et al.}{2018}]{ZDG18} Zotos E.E., Dubeibe F.L., Gonz\'{a}lez G.A., 2018, MNRAS, 477, 5388

\end{thebibliography}
\end{document}